\documentclass{jfm}
\usepackage{graphicx}
\usepackage{epstopdf, epsfig}

\shorttitle{On the Properties of Energy Flux in Wave Turbulence}
\shortauthor{A. Hrabski and Y. Pan}

\title{On the Properties of Energy Flux in Wave Turbulence}

\author{Alexander Hrabski
 \and Yulin Pan
 \corresp{\email{yulinpan@umich.edu}}}

\affiliation{Department of Naval Architecture and Marine Engineering, \\
University of Michigan, Ann Arbor, MI 48109, USA}

\begin{document}

\maketitle

\begin{abstract}
We study the properties of energy flux in wave turbulence via the Majda-McLaughlin-Tabak (MMT) equation with a quadratic dispersion relation. One of our purposes is to resolve the inter-scale energy flux $P$ in the stationary state to elucidate its distribution and scaling with spectral level. More importantly, we perform a quartet-level decomposition of $P=\sum_\Omega P_\Omega$, with each component $P_\Omega$ representing the contribution from quartet interactions with frequency mismatch $\Omega$, in order to explain the properties of $P$ as well as study the wave-turbulence closure model. Our results show that time series of $P$ closely follows a Gaussian distribution, with its standard deviation several times its mean value $\overline{P}$. This large standard deviation is shown to mainly result from the fluctuation (in time) of the quasi-resonances, i.e., $P_{\Omega\neq 0}$. The scaling of spectral level with $\overline{P}$ exhibits $\overline{P}^{1/3}$ and $\overline{P}^{1/2}$ at high and low nonlinearity, consistent with the kinetic and dynamic scalings respectively. The different scaling laws in the two regimes are explained through the dominance of quasi-resonances ($P_{\Omega\neq 0}$) and exact resonances ($P_{\Omega= 0}$) in the former and latter regimes. Finally, we investigate the wave-turbulence closure model, which connects fourth-order correlators to products of pair correlators through a broadening function $f(\Omega)$, sometimes argued to be a $sinc$ function in the theory. Our numerical data show that consistent behavior of $f(\Omega)$ can only be observed upon averaging over a large number of quartets, but with $f(\Omega)$ showing $f\sim 1/\Omega^\beta$ dependence with $\beta$ taking values between $1.3$ and $1.6$.

\end{abstract}


\section{Introduction}

Wave turbulence theory (WTT) is a framework for describing the long-time statistical behavior of fields of many weakly-interacting waves. In such a system, over scales far from those of forcing and dissipation, self-similar wave-wave interactions drive an energy cascade between scales and lead to the development of a power-law spectrum. While wave turbulence shares much phenomenology with hydrodynamic turbulence, WTT enables an analytic treatment of the governing wave equation through which the evolution equation of the wave spectrum can be derived, yielding as a stationary solution the full functional form of the power-law spectrum. Due to the ubiquity of nonlinear waves in nature, WTT has found use in a diverse array of fields, including magnetohydrodynamics \citep{galtier_weak_2000,galtier_weak_2014}, physical oceanography \citep{zakharov_weak_1967,zakharov_stability_1968}, acoustics \citep{lvov_statistical_1997}, astrophysics \citep{galtier_direct_2021}, and others.

One of the primary results of WTT is the derivation of the Wave Kinetic Equation (WKE), which expresses the time evolution of wave action spectral density as an integral over wave-wave interactions. The WKE has an inertial-range stationary solution of wave action $n(k)\sim P^{\theta}k^{\gamma}$, where $k$ is the wave number, $P$ is the energy flux of the forward cascade, and $\theta$ and $\gamma$ are scaling exponents. Over the decades, many efforts have been made to numerically and experimentally study the scaling exponent $\gamma$ for a wide variety of physical systems \cite[e.g.][]{nazarenko_wave_2006,denissenko_gravity_2007,miquel_role_2014,during_wave_2017,hassaini_confinement_2018,monsalve_quantitative_2020}. The exponent $\theta$, and in general the properties of $P$, are however much less studied, despite the fact that they are more relevant to the formulation of the WKE. A small number of existing works on the scaling of $P$ \cite[e.g.][]{falcon_observation_2007,deike_energy_2014,pan_direct_2014} sometimes produce inconsistent conclusions, partly because of their indirect and inconsistent measurements of $P$ based on the energy input/dissipation rate that are inevitably complicated by the non-stationary spectrum and broadscale dissipation  of the wave field \cite[e.g.][]{deike_energy_2014,pan_decaying_2015}.

A more direct approach for exact evaluation of $P$ can be formulated through the nonlinear terms in the governing equation which are responsible for the wave-wave interactions leading to the energy cascade \cite[e.g.][]{hrabski_effect_2020}. This approach allows the resolution of the probability distribution of $P$ at arbitrary scales that is not obtainable by previous input/dissipation-based methods. Moreover, this formulation enables a decomposition of the energy flux into contributions from exact and quasi resonances. In particular, we can perform a quartet-level decomposition of $P=\sum_{\Omega} P_{\Omega}$, with $P_{\Omega}$ representing  contributions from a set of quartets with frequency mismatch $\Omega$. This decomposition technique will allow us to elucidate many mechanisms underlying the scaling and distribution of $P$, and provide a direct measure of nonlinear broadening by evaluating the contribution of quasi-resonances to the energy cascade. This new measure of nonlinear broadening can be more direct and physically intuitive than previous approaches based on the coherence function \cite[e.g.][]{aubourg_nonlocal_2015,pan_understanding_2017,zhang_numerical_2021} or the broadened dispersion relation \cite[e.g.][]{mordant_fourier_2010,deike_direct_2014}.

Another study enabled by this decomposition technique is one on the WTT closure model, which relates the high-order correlators (of frequency mismatch $\Omega$) to the product of pair correlators. This relation is usually argued in conjunction with a broadening function $f(\Omega)$ that approaches the delta function $\delta(\Omega)$ at the kinetic limit (of a large domain and small nonlinearity) \citep{nazarenko_wave_2011,zakharov_kolmogorov_2012,buckmaster_onset_2021,deng_full_2021,deng_derivation_2021}. We are particularly interested in the form of $f(\Omega)$ in a finite domain, which is more relevant to the situations in many numerical simulations and experiments. Interesting attempts in this direction include the development of a generalized kinetic equation (by implementing a numerical solution to the closure) \citep{annenkov_role_2006}, which however does not focus on the functional form $f(\Omega)$ in the stationary state of wave turbulence. A lack of understanding of the closure problem is a major obstacle for the advancement in the theory of wave turbulence, e.g., the obvious contradiction between the MMT closure and WTT closure (as well as the elusive numerical observations) raised more than 20 years ago is still not fully understood today \citep{majda_one-dimensional_1997,cai_spectral_1999,zakharov_wave_2001}. 

The purpose of this paper, in general, is to establish a methodology such that all aforementioned analysis (including various properties of energy flux and the closely-related closure problem) can be directly studied using numerical data from simulations of the governing equations. We demonstrate our methodology in the context of the two-dimensional Majda-McLaughlin-Tabak (MMT) equation, but envision applications to much broader systems in wave turbulence. In a stationary state of MMT turbulence, our analysis shows that the energy flux $P$, as a time series, closely follows a Gaussian distribution, with a standard deviation of several times its mean value $\overline{P}$. The large standard deviation of $P$ is found to be dominated by fluctuations in time of the quasi-resonant components $P_{\Omega \neq 0}$. The scaling between the spectral level and $\overline{P}$ shows a $\overline{P}^{1/3}$ dependence at high nonlinearity levels and transits to $\overline{P}^{1/2}$ dependence at low nonlinearity levels. The $\overline{P}^{1/3}$ scaling is consistent with the kinetic scaling and is established when $\overline{P}$ is dominated by the quasi-resonant contributions. This is a remarkable result considering the fact that the WKE, based on which the kinetic scaling is developed, is formulated on the exact resonant manifold. The $\overline{P}^{1/2}$ scaling is consistent with the dynamic scaling derived from the MMT equation, established as a result of the dominance of exact resonances in $\overline{P}$. Finally, the WTT closure model on the fourth-order correlator, when evaluated in a time window of $O(500)$ fundamental periods, is found to be not valid on the individual quartet level. When we increase the number of quartets to the level that $P$ is formulated ($O(10^9)$ number of quartets for each $\Omega$), the average behavior of the fourth-order correlator lies much closer to WTT closure, but with $f(\Omega)\sim 1/\Omega^\beta$ (with $\beta$ between 1.3 and 1.6) observed instead of the previously argued functions (e.g., $sinc(\Omega)$) from WTT.

\section{Formulation of Energy Flux for the MMT Model}
We consider a two-dimensional (2D) MMT model \citep{majda_one-dimensional_1997} which has been widely used to study wave turbulence problems \citep{cai_spectral_1999,sheffield_ensemble_2017,chibbaro_weak_2017,hrabski_effect_2020}. The model describes the evolution of a complex scalar $\psi(\boldsymbol{x},t)$:
\begin{equation}
     i\frac{\partial \psi}{\partial t} = |\partial_{\boldsymbol{x}}|^\alpha\psi + |\partial_{\boldsymbol{x}}|\left(\big| |\partial_{\boldsymbol{x}}|\psi\big|^2|\partial_{\boldsymbol{x}}|\psi\right),
     \label{eqn:MMT}
\end{equation}
where $\boldsymbol{x}$ is the 2D spatial coordinates and $t$ is time, and the operator $|\partial_{\boldsymbol{x}}|^{\alpha}\psi$ corresponds to the multiplication of each Fourier component $\hat{\psi}_{\boldsymbol{k}}$ by $k^{\alpha}$ with $k$ being the magnitude of wave number vector $\boldsymbol{k}$. We choose $\alpha=2$, yielding a dispersion relation $\omega_{k}=k^2$ that is the same as the nonlinear Schrodinger equation. 

\subsection{Exact Formulations of Instantaneous Energy Flux $P(t)$}
To formulate the energy flux $P$ across arbitrary wave number $k_b$, we start from a consideration of energy conservation in a control volume $k<k_b$ in spectral space:
\begin{equation}
    P(t) = -\sum_{ \boldsymbol{k}\in \{ \boldsymbol{k}|k<k_b \} }\omega_{k}\frac{\partial(\hat{\psi}_{\boldsymbol{k}}\hat{\psi}_{\boldsymbol{k}}^*)}{\partial t}(t) = - \sum_{ \boldsymbol{k}\in \{ \boldsymbol{k}|k<k_b \} }\omega_{k}\left( \frac{\partial\hat{\psi}_{\boldsymbol{k}}}{\partial t}\hat{\psi}_{\boldsymbol{k}}^* +  \frac{\partial \hat{\psi}_{\boldsymbol{k}}^*}{\partial t} \hat{\psi}_{\boldsymbol{k}}\right)(t).
    \label{eqn:ptot}
\end{equation}
Substituting  the Fourier-domain representation of (\ref{eqn:MMT}) into (\ref{eqn:ptot}) yields
\begin{equation}
    P(t) = -\sum_{ \boldsymbol{k}\in \{ \boldsymbol{k}|k<k_b \} }\omega_{k} \sum_{(\boldsymbol{k}_{1}, \boldsymbol{k}_{2}, \boldsymbol{k}_{3})\in S_{\boldsymbol{k}}} 2 k_1 k_2 k_3 k \Imag\left(\hat{\psi}_{\boldsymbol{1}} \hat{\psi}_{\boldsymbol{2}} \hat{\psi}_{\boldsymbol{3}}^* \hat{\psi}_{\boldsymbol{k}}^*\right)(t),
    \label{eqn:pquartet}
\end{equation}
where $S_{\boldsymbol{k}}$ is the set of all $(\boldsymbol{k}_{1}, \boldsymbol{k}_{2}, \boldsymbol{k}_{3})$ with $\boldsymbol{k}_{1} + \boldsymbol{k}_{2} - \boldsymbol{k}_{3} - \boldsymbol{k}=0$. We note that (\ref{eqn:pquartet}) is exact even if forcing and dissipation are added to (\ref{eqn:MMT}) (which may affect (\ref{eqn:ptot}) but not (\ref{eqn:pquartet})), because only the nonlinear term in (\ref{eqn:MMT}) is responsible for the inter-scale energy flux. Taking the wave action $n({\boldsymbol{k}})\sim \hat{\psi}_{\boldsymbol{k}}^2$, (\ref{eqn:pquartet}) implies a dynamic scaling of $n\sim P^{1/2}$ via a heuristic argument. This is a result of directly using the dynamic equation (\ref{eqn:MMT}) to formulate the energy flux.

With (\ref{eqn:pquartet}) available, we can further perform a decomposition $P(t)=\sum_{\Omega} P_{\Omega}(t)$ by partitioning the set $S_{\boldsymbol{k}}$ according to the frequency mismatch of each quartet interaction. Specifically, defining $S_{\Omega , \boldsymbol{k}} \equiv\ \{ (\boldsymbol{k}_1, \boldsymbol{k}_2, \boldsymbol{k}_3) \in S_{\boldsymbol{k}} \ \big| \  |\omega_{1}+\omega_{2}-\omega_{3}-\omega_{k}| = \Omega  \}$, we have $\bigcup_{\Omega} S_{\Omega ,\boldsymbol{k}} = S_{\boldsymbol{k}}$ with all sets $S_{\Omega ,\boldsymbol{k}}$ as disjoint. Therefore, $P_{\Omega}(t)$ can be naturally formulated as 
\begin{equation}
    P_\Omega(t) = -\sum_{ \boldsymbol{k}\in \{ \boldsymbol{k}|k<k_b \} } \omega_{k} \sum_{(\boldsymbol{k}_{1}, \boldsymbol{k}_{2}, \boldsymbol{k}_{3})\in S_{\Omega , \boldsymbol{k}}} 2 k_1 k_2 k_3 k \Imag\left(\hat{\psi}_{\boldsymbol{1}} \hat{\psi}_{\boldsymbol{2}} \hat{\psi}_{\boldsymbol{3}}^* \hat{\psi}_{\boldsymbol{k}}^*\right)(t).
    \label{eqn:Pex}
\end{equation}
The computation using (\ref{eqn:Pex}) allows us to measure the contribution to $P$ from resonances with different $\Omega$, as well as to separate the quasi-resonant and exact-resonant contributions by $P_{\Omega>0}$ and $P_{\Omega=0}$, respectively.

\subsection{Formulations of Energy Flux under the WTT Closure}
The WTT closure for (\ref{eqn:MMT}) relates the fourth-order correlator to pair correlators by 
\begin{eqnarray}
    \Imag \overline{\left(\hat{\psi}_{\boldsymbol{1}}\hat{\psi}_{\boldsymbol{2}}\hat{\psi}_{\boldsymbol{3}}^{*}\hat{\psi}_{\boldsymbol{k}}^{*}\right)} _{\Omega}
    = 2 k k_1 k_2 k_3 (n_{\boldsymbol{1}} n_{\boldsymbol{2}} n_{\boldsymbol{3}} + n_{\boldsymbol{1}} n_{\boldsymbol{2}} n_{\boldsymbol{k}} - n_{\boldsymbol{1}} n_{\boldsymbol{k}} n_{\boldsymbol{3}} - n_{\boldsymbol{k}} n_{\boldsymbol{2}} n_{\boldsymbol{3}})f(\Omega),
    \label{eqn:closure}
\end{eqnarray}
where an over bar denotes the ensemble average (or time average in numerical analysis) and $n_{\boldsymbol{k}} = \overline{\hat{\psi}_{\boldsymbol{k}} \hat{\psi}^*_{\boldsymbol{k}}}$ is wave action. We also use a subscript $\Omega$ for $\Imag \overline{\left(\hat{\psi}_{\boldsymbol{1}}\hat{\psi}_{\boldsymbol{2}}\hat{\psi}_{\boldsymbol{3}}^{*}\hat{\psi}_{\boldsymbol{k}}^{*}\right)} _{\Omega}$ to denote the frequency mismatch of the corresponding four wave modes. The closure (\ref{eqn:closure}) has been derived in various (heuristic) ways in the physics literature \citep{zakharov_kolmogorov_2012,janssen_nonlinear_2003,nazarenko_wave_2011,pan_understanding_2017-1}, all assuming quasi-Gaussian statistics and quasi-stationary spectra (to obtain an analytical solution of the differential equation for fourth-order correlators). Depending on different methods of derivation, $f(\Omega)$ takes the form of $\sin(\Omega t)/\Omega$ \citep{janssen_nonlinear_2003} or $\epsilon/(\Omega^2+\epsilon^2)$ \citep{zakharov_kolmogorov_2012}. The WKE is then derived in the limit of the parameters $t\rightarrow\infty$ and $\epsilon\rightarrow0$ for the first and second cases respectively, with both leading to $f(\Omega)\rightarrow \pi\delta(\Omega)$.

We are interested in the performance of the closure (\ref{eqn:closure}) in computing the energy flux $P_\Omega$. For this purpose, we substitute (\ref{eqn:closure}) into the average of (\ref{eqn:Pex}) to obtain
\begin{eqnarray}
      \overline{P}_{\Omega} = -\sum_{ \boldsymbol{k}\in \{ \boldsymbol{k}|k<k_b \} } \omega_{k} & \nonumber \\ 
        \times \sum_{\boldsymbol{k}_{1}, \boldsymbol{k}_{2}, \boldsymbol{k}_{3}\in S_{\Omega , \boldsymbol{k}}} 4 k_1^2 k_2^2 k_3^2 k^2 & (n_{\boldsymbol{1}} n_{\boldsymbol{2}} n_{\boldsymbol{3}} + n_{\boldsymbol{1}} n_{\boldsymbol{2}} n_{\boldsymbol{k}} - n_{\boldsymbol{1}} n_{\boldsymbol{k}} n_{\boldsymbol{3}} - n_{\boldsymbol{k}} n_{\boldsymbol{2}} n_{\boldsymbol{3}} )f(\Omega).
    \label{eqn:Pke2}
\end{eqnarray}

In \S 4, we will evaluate the functional form $f(\Omega)$ in both (\ref{eqn:closure}) and (\ref{eqn:Pke2}) using our numerical data to compute all other terms in the two equations (e.g., $\overline{P}_{\Omega}$ in (\ref{eqn:Pke2}) can be computed through time average of (\ref{eqn:Pex})). The difference between the two evaluations of $f(\Omega)$ is that the latter involves the sum of an enormous number of interactions while the former is for an individual quartet. Through comparison of the numerically resolved $f(\Omega)$ with their WTT counterparts, the validity of the WTT closure can be assessed. 

\section{Setup of Numerical Experiments} \label{sec:numer}
We compute the solution to (\ref{eqn:MMT}) via a pseudospectral method on a periodic domain of size $2\upi \times 2\upi$ containing $512\times512$ modes. The linear term is integrated analytically to reduce the system stiffness, while the nonlinear term is integrated via an explicit fourth order Runge-Kutta scheme. Our purpose is to generate a long stationary state so that the distributions of $P$ (and other quantities discussed in \S 2) are sufficiently resolved. Therefore, we force the system at large scales (in conjunction with small-scale dissipation) instead of considering free-decay turbulence. Specifically, we add a forcing term
\begin{equation}
    F = \left\{ \begin{array}{l}
    F_{r}+iF_{i} \ \ \ 7\le k\le9 \\
    0 \ \ \ otherwise,
    \end{array} \right.
    \label{eqn:forcing}
\end{equation}
to the right hand side of (\ref{eqn:MMT}), with $F_{r}$ and $F_{i}$ taken from a Gaussian distribution with zero mean and a variance $\sigma^2_F$ (which determines the forcing magnitude). Dissipation is accounted for by adding two terms
\begin{eqnarray}
    D_1 =& \left\{ \begin{array}{l}
    -i\nu_1 \hat{\psi}_{\boldsymbol{k}} \ \ \ k\ge100 \\
    0 \ \ \ otherwise,
    \end{array} \right. \nonumber \\
    D_2 =& \left\{ \begin{array}{l}
    -i\nu_2 \hat{\psi}_{\boldsymbol{k}} \ \ \ k\le7 \\
    0 \ \ \ otherwise,
    \end{array} \right.
    \label{eqn:dissipation}
\end{eqnarray}
at small and large scales respectively, with the latter included to prevent energy accumulation at large scales due to the inverse cascade. The parameters in (\ref{eqn:dissipation}) are chosen to be $\nu_1 = 6\times10^{-12} (k-100)^{8}$ and $\nu_2 = 30 k^{-4}$ throughout all the simulations. Forcing and dissipation of this type have been demonstrated to produce results compatible with the WTT predictions in the one-dimensional MMT model \citep{cai_spectral_1999}.

In addition, to accelerate convergence to the stationary state, we start the simulations from initial conditions described by Gaussian spectra $\hat{\psi}_{\boldsymbol{k}}=a_0e^{-0.1|k-10|+\mathrm{i}\phi_{\boldsymbol{k}}}$, with $\phi_{\boldsymbol{k}}$ the uniformly-distributed, decorrelated random phases, and $a_0$ a real constant chosen to provide an energy close to that of the expected stationary state. To obtain the scaling $\theta$ over a range of $\overline{P}$, we run a collection of 19 simulations, differing only in forcing strength $\sigma^2_F$ and initial spectral level $a_0$. These simulations cover a range of $\overline{P}$ spanning several orders of magnitude, with data collected in the stationary state for each case.

\section{Results} \label{sec:res}
\begin{figure}
  \centerline{\includegraphics{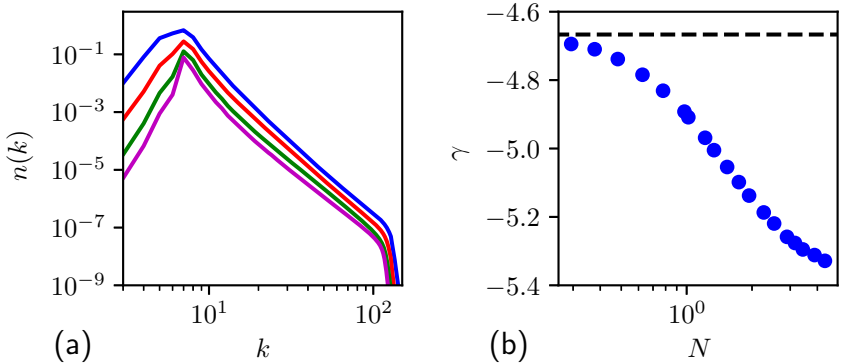}}% Images in 100% size
  \caption{(a) A representative collection of fully-developed, angle-averaged wave action spectra $n(k)$. (b) Spectral slope $\gamma$ as a function of $N$, with WTT value $\gamma_0=-4.67$ indicated (\dashed).}
\label{fig:niso}
\end{figure}

Before presenting the results on energy flux, we first check the spectra at stationary states in simulations with different forcing magnitudes. Several typical spectra at different levels are shown in figure \ref{fig:niso}a, where we observe power-law ranges close to one decade for all spectra. Figure \ref{fig:niso}b shows the power-law exponent $\gamma$ evaluated in all 19 simulations, as a function of the spectral level computed by an integral measure of the (conservatively taken) power-law range
\begin{equation}
    N = \sum_{ \boldsymbol{k}\in \{ \boldsymbol{k}|13<k<60 \} } n_{\boldsymbol{k}}.
    \label{eqn:N}
\end{equation}
We see in figure \ref{fig:niso}b that $\gamma$ increases (i.e., the spectrum becomes shallower) with the decrease of $N$, reaching the WTT prediction $\gamma_0=-4.67$ for low spectral levels. This behavior of $\gamma$ is consistent with our previous study of the free-decay MMT turbulence (except that use of different dissipation schemes may have some slight effect). As analyzed by the authors \citet{hrabski_effect_2020}, the fact that $\gamma\rightarrow \gamma_0$ for small $N$ is a result of the dispersion relation $\omega_k=k^2$ which leads to a continuous resonant system at low nonlinearity \citep{faou_weakly_2016}. The deviation of $\gamma$ from $\gamma_0$ at high nonlinearity may result from coherent structures, as suggested by \citet{zakharov_wave_2001} and \cite{chibbaro_weak_2017} in the one-dimensional context, or some features of the 2D MMT model that are yet to be fully understood. For waves in different physical contexts, e.g., surface gravity waves \citep{zhang_numerical_2021} and capillary waves \citep{pan_direct_2014}, the behaviors of $\gamma$ are remarkably different.

We next present our full study of energy flux, with results organized into three sections. \S 4.1 discusses the distributions of $P$ and its associated decomposition $P_\Omega$. \S 4.2 focuses on the scaling of spectral level with $P$, with the results explained by the contributions of quasi/exact resonances to $P$. The study related to the closure model is then presented in \S 4.3. 

\subsection{Flux Distributions and Decomposition}
\begin{figure}
  \centerline{\includegraphics{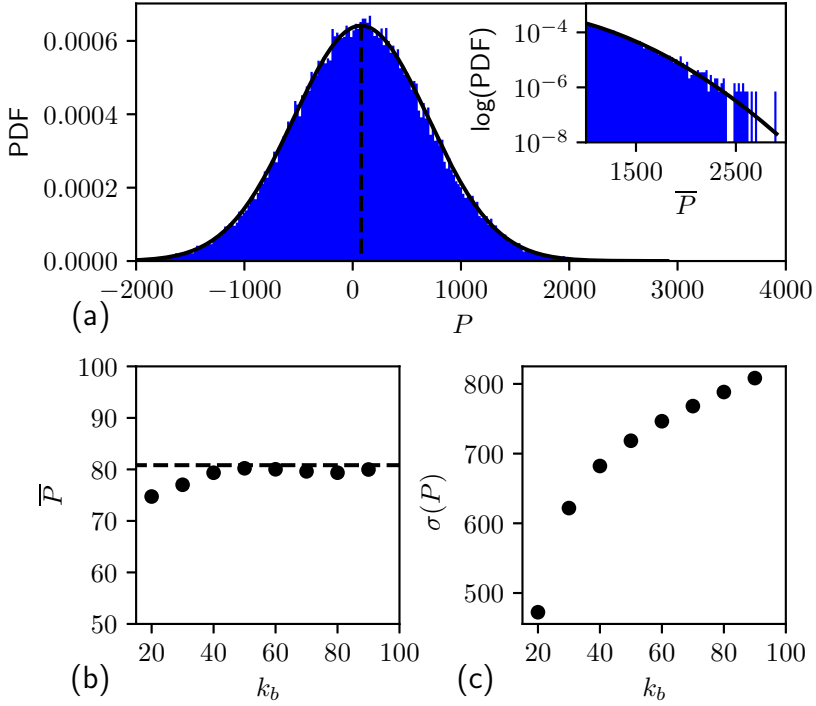}}% Images in 100% size
  \caption{(a) The histogram of stationary time series $P(t)$ evaluated over $256 T_0$, fitted with a Gaussian distribution of the same mean and standard deviation (\full) for reference. Figure inset: tail of the distribution in logarithmic scale. (b) the mean $\overline{P}$ and (c) standard deviation $\sigma(P)$ evaluated for different $k_b$. The dissipation-based estimate of $P_d$ is indicated in both (a) and (b) by (\dashed).}
\label{fig:pk}
\end{figure}

A typical distribution of energy flux $P(t)$, computed with $k_b=30$ from $2^{16}$ data points over a time window of $T_w=256T_0$ (with $T_0$ the fundamental period), is shown in figure \ref{fig:pk}a. We find that $P$ closely follows a Gaussian distribution, with a standard deviation $\sigma(P)=621.8$ several times larger than the mean value $\overline{P}=77.02$. The very large standard deviation is consistent with previous studies in wave turbulence \citep{falcon_fluctuations_2008} and hydrodynamic turbulence \citep{bandi_energy_2006}. However, the nearly perfect Gaussian form of the distribution has not been observed in these previous works, probably because of their approximation in evaluating $P$ (either from energy input rate or a filter-based method) and the different turbulent systems considered in their study. 
In addition, our method allows us to study the fully-resolved distribution of $P$ at any scale (i.e., with arbitrary $k_b$). In figures \ref{fig:pk}b and \ref{fig:pk}c, we plot the values of $\overline{P}$ and $\sigma(P)$ for $k_b$ varying in the inertial range from 20 to 90. The mean flux $\overline{P}$ remains almost constant for all $k_b$, which is consistent with the WTT constant flux argument in the inertial range (this is only possible by avoiding broad-scale dissipation in simulations). The standard deviation $\sigma(P)$ increases with $k_b$, because more quartet interactions are included (as $k$ becomes denser) resulting in more fluctuations in $P(t)$. We also include in figures \ref{fig:pk}a and \ref{fig:pk}b the energy flux $\overline{P}$ computed from the high-wave-number dissipation rate
\begin{equation}
    \overline{P}_d = \sum_{ \boldsymbol{k}\in \{ \boldsymbol{k}|k>100 \} }\nu_1 \omega_{k} \overline{\hat{\psi}_{\boldsymbol{k}} \hat{\psi}_{\boldsymbol{k}}^*},
    \label{eqn:pd}
\end{equation}
which agrees well with the majority values of $\overline{P}$, especially for larger $k_b$ (to a relative difference within $O(1\%)$). Because of this, we will use $\overline{P}_d$ to represent the values of $\overline{P}$ for all 19 simulations in the subsequent analysis, since $P_d$ yields a faster calculation due to an easier formulation and much smaller fluctuations (requiring less data points for averaging).
\begin{figure}
  \centerline{\includegraphics{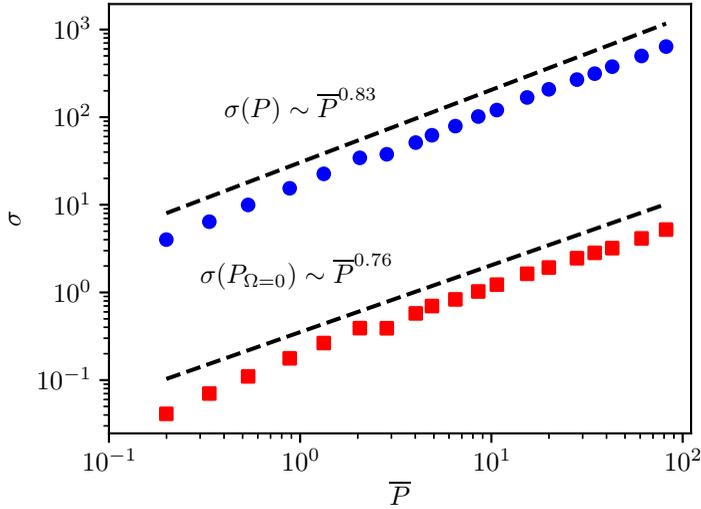}}% Images in 100% size
  \caption{Dependence of $\sigma(P)$ (\fullcirc) and $\sigma(P_{\Omega=0})$ (\fullsquare) on $\overline{P}$, with the best fits indicated (\dashed).}
\label{fig:pscale}
\end{figure}
We next examine the relation between $\sigma(P)$ and nonlinearity level measured by $\overline{P}$, with $\sigma(P)$ as a function of $\overline{P}$ plotted in figure \ref{fig:pscale}. The result shows a power-law relation over two decades given by $\sigma(P) \sim \overline{P}^{0.8 \pm 0.05}$. Furthermore, we include in figure \ref{fig:pscale} the standard deviation of the exact-resonant contributions to energy flux, $\sigma(P_{\Omega=0})$, with $P_{\Omega=0}$ calculated by the decomposition method presented in \S 2. We observe a similar power-law relation between  $\sigma(P_{\Omega=0})$ and $\sigma(P)$, but with the value of $\sigma(P_{\Omega=0})$ $O(10)$ times smaller than $\sigma(P)$ consistently for each nonlinearity level. This indicates that the large fluctuations in $P(t)$ are mainly generated due to quasi-resonant interactions. 

\begin{figure}
  \centerline{\includegraphics{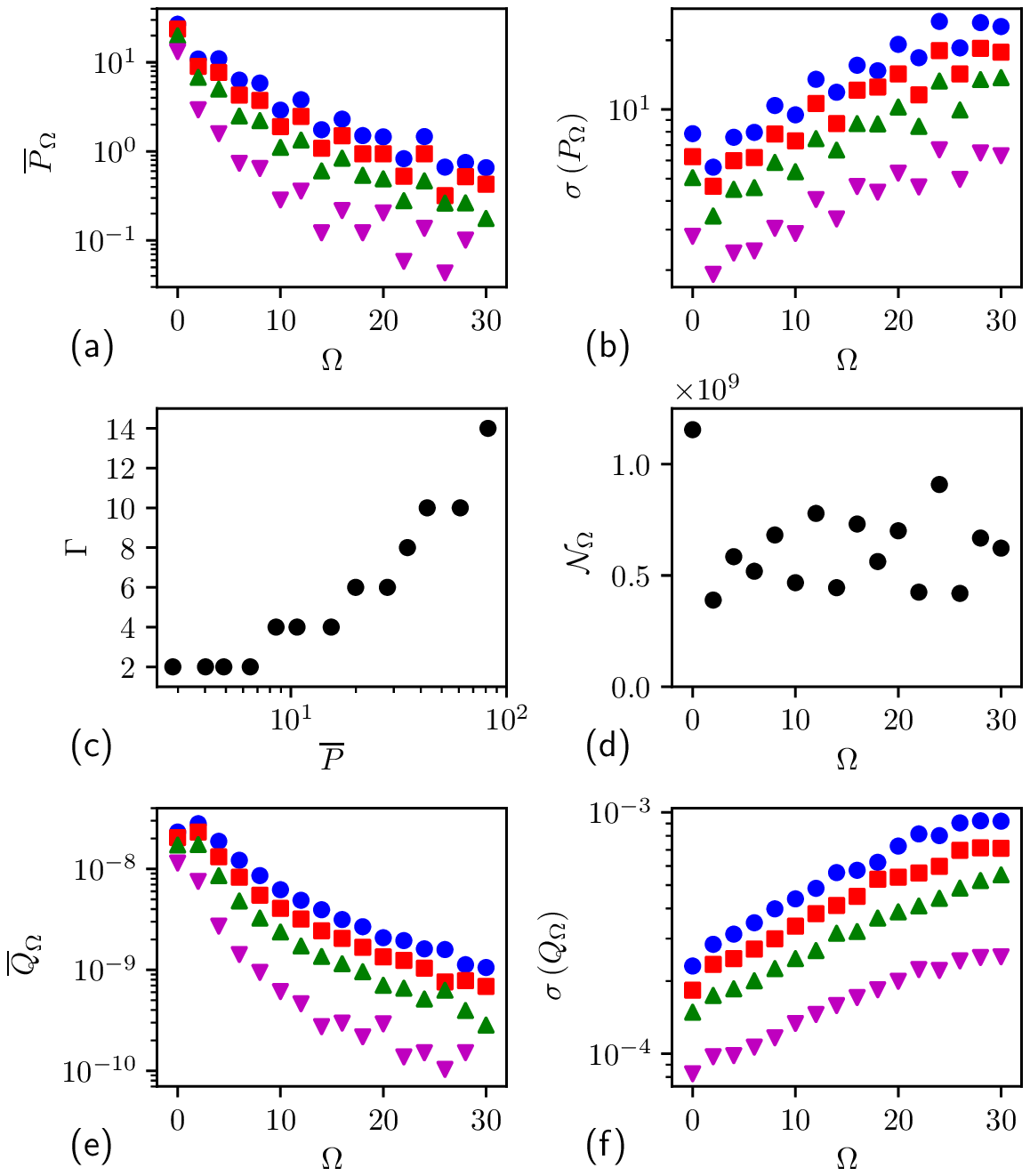}}% Images in 100% size
  \caption{(a) $\overline{P}_\Omega$ and (b) $\sigma(P_\Omega)$ as functions of $\Omega$, for four levels of nonlinearity with $\overline{P}=81.9$ (\fullcirc), $61.0$ (\fullsquare), $42.8$ (\fulltri), and $20.0$ (\fulltriangledown); (c) nonlinear broadening $\Gamma$ as a function of $\overline{P}$; (d) number of quartet interactions $\mathcal{N}_{\Omega}$ for different $\Omega$; (e) and (f) are similar to (a) and (b) but plotted for normalized flux $Q_n$. The computations to generate these results are conducted for $k_b=23$ to reduce the computational cost associated with the number of involved quartets.} 
\label{fig:pdw1}
\end{figure}

A more detailed study about the contributions of exact and quasi resonances to $\overline{P}$ and $\sigma(P)$ can be conducted by looking into the components of $P_\Omega$ for varying values of $\Omega$. In figures \ref{fig:pdw1}a and \ref{fig:pdw1}b we plot $\overline{P}_\Omega$ and $\sigma(P_\Omega)$ for $\Omega\in [0,30]$ at four different levels of nonlinearity. We note that $\Omega$ can only take even integer values for the dispersion relation $\omega_k=k^2$ on a periodic domain of $2\upi\times2\upi$. The general trends in  figure \ref{fig:pdw1}a and \ref{fig:pdw1}b show that $\overline{P}_\Omega$ decreases, but $\sigma(P_\Omega)$ increases with the increase of $\Omega$. This corresponds to a physical picture that as the interactions become more ``quasi'' (i.e., frequency mismatch $\Omega$ becomes larger), they contribute less to the mean flux but may contribute more to the fluctuations of the flux. We also emphasize here that while we always have $\sum_\Omega \overline{P}_\Omega = \overline{P}$, the quantity $\sum_\Omega \sigma^2(P_\Omega)$ is in general not equal to $\sigma^2(P)$ because $P_\Omega(t)$ with different $\Omega$ are not independent. Nevertheless, figure \ref{fig:pdw1}b in conjunction with figure \ref{fig:pscale} are sufficient to support the dominance of quasi-resonances in generating the large fluctuations in $P(t)$.

We conclude this section by summarizing two additional important results regarding $P_\Omega$. First, the decomposition in terms of $\Omega$ enables a direct measure of nonlinear broadening by quantitatively considering the contribution of quasi-resonances to the total energy flux. For example, the nonlinear broadening $\Gamma$ based on the definition $\Gamma \equiv min \{\Omega|\overline{P}_{\Omega}< 0.1\times \overline{P}_{\Omega=0} \}$ is plotted in figure \ref{fig:pdw1}c, showing that $\Gamma$ increases with $\overline{P}$ (or the nonlinearity level). Second, the fluctuations seen in figure \ref{fig:pdw1}a and \ref{fig:pdw1}b can be removed by considering the normalized flux $Q_\Omega(t)=P_\Omega(t)/\mathcal{N}_\Omega$. With $\mathcal{N}_\Omega$ (shown in figure \ref{fig:pdw1}d) counting the number of elements in $\sum_{ \boldsymbol{k}\in \{ \boldsymbol{k}|k<k_b \} } S_{\Omega , \boldsymbol{k}}$ (see (\ref{eqn:Pex})), $Q_\Omega(t)$ calculates the quartet-averaged flux (over quartets with frequency mismatch $\Omega$). As plotted in figure \ref{fig:pdw1}e and \ref{fig:pdw1}f, both $\overline{Q}_{\Omega}$ and $\sigma(Q_\Omega)$ behave smoothly for the range of $\Omega$. We speculate that these smooth behaviors regarding $Q_\Omega$ may hold for different domain geometries (e.g., tori with different aspect ratios) with different functions $\mathcal{N}_\Omega$.

\subsection{Scaling of Spectral Level with Flux}
\begin{figure}
  \centerline{\includegraphics{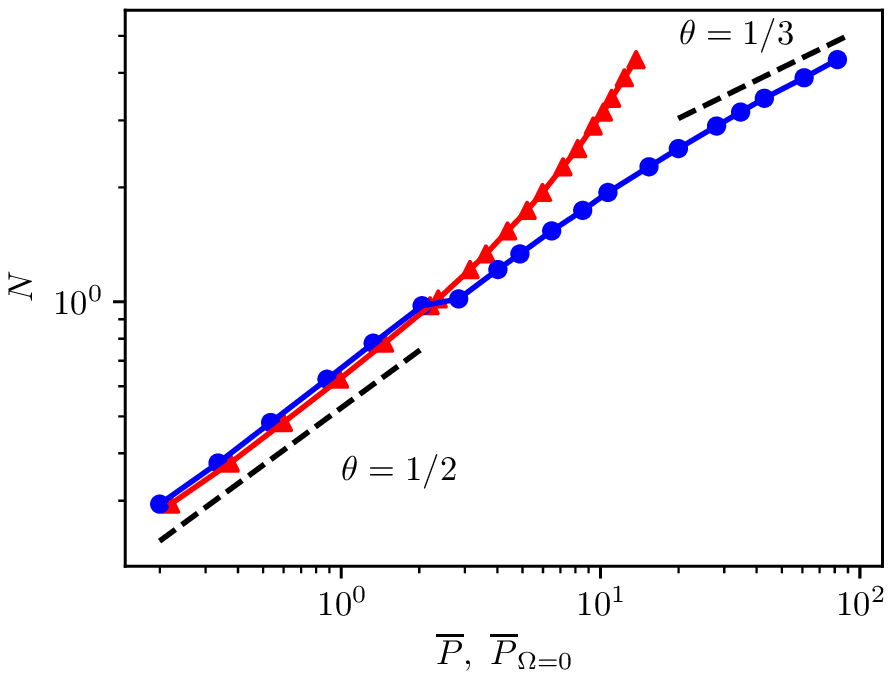}}% Images in 100% size
  \caption{The scaling of inertial-range wave action $N$ with $\overline{P}$ (\fullcircline) and $\overline{P}_{\Omega =0}$ (\fulltriline). The dynamic scaling $\theta=1/2$ and kinetic scaling $\theta=1/3$ are indicated (\dashed).}
\label{fig:ep}
\end{figure}

To understand the scaling of spectral level with energy flux, we plot in figure \ref{fig:ep} the spectral level $N$ (see (\ref{eqn:N})) as a function of both $\overline{P}$ and $\overline{P}_{\Omega=0}$ representing total and exact-resonant flux respectively. Two salient scalings are observed over the 3 decades of energy flux. At high nonlinearity level with $\overline{P} \in [30,100]$, we find a scaling approaching $N\sim \overline{P}^{1/3}$ (i.e., $\theta=1/3$ with $\theta$ the scaling exponent defined in \S 2) consistent with the kinetic scaling of WTT. At low nonlinearity level with $\overline{P} \in [0.3,3]$, the scaling behaves as $N\sim \overline{P}^{1/2}$ (i.e., $\theta=1/2$) consistent with the dynamic scaling from (\ref{eqn:pquartet}). We next discuss the mechanisms underlying these two scalings.

For high nonlinearity, we see in figure \ref{fig:ep} that we have $\overline{P} \gg \overline{P}_{\Omega=0}$ consistent with observations in figure \ref{fig:pdw1}a. This suggests that the kinetic scaling, although developed from the WKE formulated on the exact resonant manifold, relies on the dominant contribution from quasi-resonances to be realized in a finite domain. This physical picture is consistent with the numerical study by \citet{annenkov_role_2006}, as well as the recent mathematical justification of the WKE \citep{deng_full_2021,deng_derivation_2021,buckmaster_onset_2021}. In particular, the rigorous mathematical derivation proves that the WKE describes the effective dynamics of both exact \emph{and} quasi resonances in the kinetic limit. The present results are supplemental to these previous works, which only deal with initial spectral evolution up to the kinetic time scale. Our study additionally suggests that this physical understanding holds for the long-time stationary state.

For low nonlinearity, $\overline{P}\approx \overline{P}_{\Omega=0}$ (as shown in figure \ref{fig:ep}) due to the elimination of quasi-resonant contributions to $\overline{P}$. This behavior has been rigorously proven by \citet{faou_weakly_2016} and numerically observed by \citet{hrabski_effect_2020}. Under this situation, the number of interactions contributing to $\overline{P}$ is substantially reduced, and the dynamic scaling from (\ref{eqn:pquartet}) becomes dominant. We further remark that the presence of an energy cascade, instead of frozen turbulence at low nonlinearity, is due to the dispersion relation $\omega_k=k^2$ leading to a continuous resonant system \citep{faou_weakly_2016} at $N\rightarrow 0$.

Finally, we note that the transition to the dynamic scaling from high to low nonlinearity seems to involve a small and non-smooth jump, as can be seen in figure \ref{fig:ep} (which is also observable in earlier figure \ref{fig:pscale}). This may suggest that the removal of quasi-resonant contributions is not smooth with the decrease of nonlinearity level, i.e., there exists a threshold below which the quasi-resonances are turned off non-incrementally. However, we will avoid overreaching with this point in our current work and consider it a hypothesis that requires further study. 

\subsection{Investigation on the Closure Model}
In this section, we use our numerical data to study the WTT closure model, in particular the magnitude and functional form of $f(\Omega)$. We focus this study on the nonlinearity level associated with the kinetic scaling of $\overline{P}$, for which the WTT closure is developed. Theoretically one expects $f(\Omega)$ to take the form of either $\sin(\Omega t)/\Omega$ \citep{janssen_nonlinear_2003} or $\epsilon/(\Omega^2+\epsilon^2)$ \citep{zakharov_kolmogorov_2012}, with $\int_\Omega f(\Omega)d\Omega \sim O(1)$ since both forms are generalized delta functions. The numerical evaluation of $f(\Omega)$ will be performed at: (1) an individual quartet level using (\ref{eqn:closure}); (2) a family of quartets level using (\ref{eqn:closure}) in an average manner that will be introduced shortly; and (3) an inter-scale energy flux level using (\ref{eqn:Pke2}). The general procedure is to compute $f(\Omega)$ from (\ref{eqn:closure}) and (\ref{eqn:Pke2}) with all other terms determined from the numerical data. To distinguish these computations, we denote the numerically obtained $f(\Omega)$ from the three levels respectively as $f_Q(\Omega)$, $f_F(\Omega)$ and $f_P(\Omega)$.

\begin{figure}
  \centerline{\includegraphics{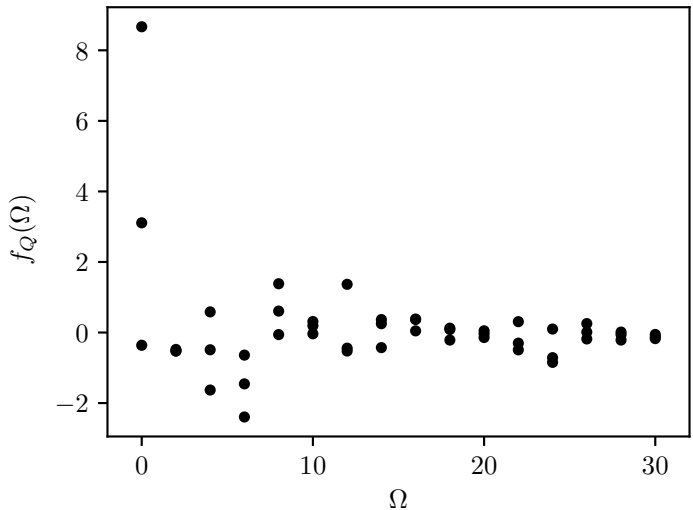}}% Images in 100% size
  \caption{The function $f_Q(\Omega)$ evaluated for $O(50)$ selected quartets (with 3 quartets for each $\Omega$) with $T_w = 256 T_0$, for the highest nonlinearity level with $\overline{P}=81.9$.}
\label{fig:q1}
\end{figure}

Figure \ref{fig:q1} shows $f_Q(\Omega)$ evaluated for $O(50)$ quartets with $\Omega\in[0,30]$, and with average quantities in (\ref{eqn:closure}) evaluated over a time window $T_w=256 T_0$. It is clear that no obvious functional pattern can be found for $f_Q(\Omega)$ (i.e., with different values of $f_Q(\Omega)$ obtained for the same $\Omega$). Also, we see some values of $f_Q(\Omega)\sim O(10)$, likely in disagreement with the WTT result of $\int_\Omega f(\Omega) d\Omega \sim O(1)$. In general, figure \ref{fig:q1} indicates that the WTT closure for fourth-order correlators cannot be used to describe the behavior of a single quartet, regardless of its associated frequency mismatch $\Omega$. 
%8477 quartets

\begin{figure}
  \centerline{\includegraphics{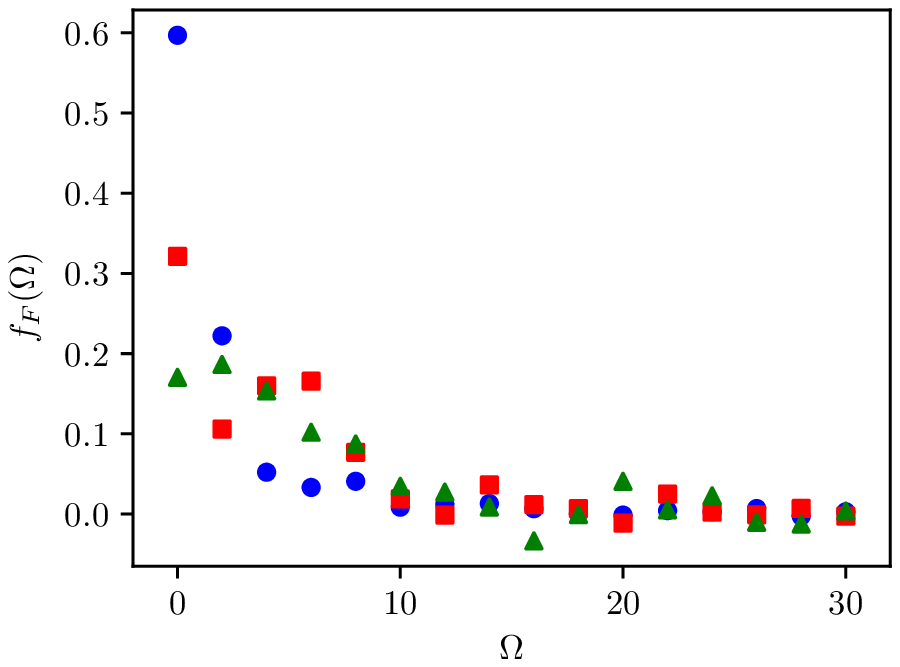}}% Images in 100% size
  \caption{The function $f_F(\Omega)$ for three representative families of quartets defined by $\boldsymbol{k}_0^e=(-2,8)$, $\boldsymbol{k}_1^e=(-10,0)$, $\boldsymbol{k}_2^e=(14+4j,-8-4j)$, and $\boldsymbol{k}_3^e=(6+4j,-16-4j)$ for $j=0$ (\fullcirc), $j=1$ (\fullsquare), and $j=2$ (\fulltri). The evaluation is for the highest nonlinearity case of $\overline{P}=81.9$ with $T_w = 256 T_0$.}
\label{fig:q2}
\end{figure}

To consider the average behavior of a family of quartets, we use a similar technique to that of \citet{annenkov_role_2006} to create a cluster of modes around each of the four modes of an exact resonant quartet. Specifically, for an exact quartet ($\boldsymbol{k}_0^e$, $\boldsymbol{k}_1^e$, $\boldsymbol{k}_2^e$, $\boldsymbol{k}_3^e$), we construct a family of quartets ($\boldsymbol{l}_0$, $\boldsymbol{l}_1$, $\boldsymbol{l}_2$, $\boldsymbol{l}_3$) at its vicinity by choosing all $\boldsymbol{l}_i$ with $|\boldsymbol{l}_i- \boldsymbol{k}_i^e|\le4$ for $i=0,1,2,3$. We then evaluate $f_F(\Omega)$ from (\ref{eqn:closure}) by summing over those quartets ($\boldsymbol{l}_0$, $\boldsymbol{l}_1$, $\boldsymbol{l}_2$, $\boldsymbol{l}_3$) with frequency mismatch $\Omega$ on both sides of the equation. Under this evaluation $f_F(\Omega)$ reflects the closure behavior averaged over $O(10^3)$ quartets. Figure \ref{fig:q2} shows $f_F(\Omega)$ computed from three representative families of quartets, defined via exact quartets ($\boldsymbol{k}_0^e$, $\boldsymbol{k}_1^e$, $\boldsymbol{k}_2^e$, $\boldsymbol{k}_3^e$). We see that $f_F(\Omega)$ is somewhat inversely proportional to $\Omega$ superposed with many fluctuations. While the general trend of $f_F(\Omega)$ seems consistent for different families, the details (e.g, the value of $f_F(0)$ as well as the fluctuation patterns) vary considerably across different families. 

\begin{figure}
  \centerline{\includegraphics{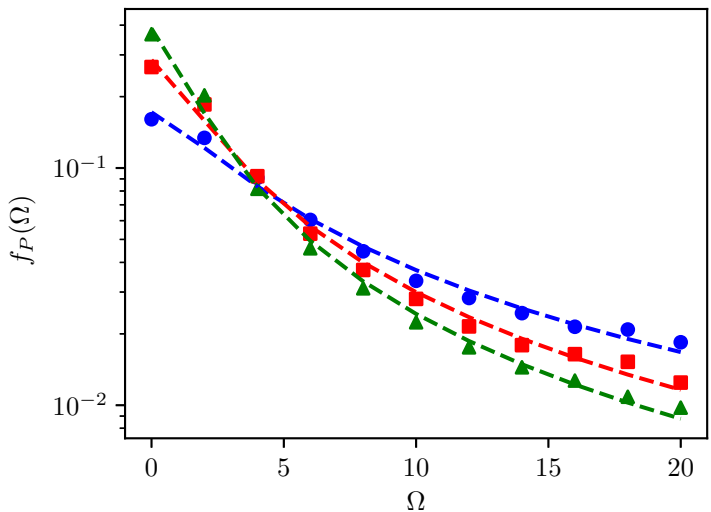}}% Images in 100% size
  \caption{The function $f_P(\Omega)$ evaluated by (\ref{eqn:Pke2}) with $k_b=23$ and $T_w = 256 T_0$, for different nonlinearity levels with $\overline{P}=81.9$ (\fullcirc), $42.8$ (\fullsquare), and $28.0$ (\fulltri). Fits to the data of the form $f(\Omega)=C/(\rho + \Omega^{\beta})$ are indicated (\dashed).}
\label{fig:f}
\end{figure}

We finally examine the closure behavior considering the average over an enormous number of quartets, chosen as all quartets contributing to the energy flux across $k_b=23$. Under such a consideration, both sides of (\ref{eqn:Pke2}) are subject to summation over $O(10^9)$ elements in $\sum_{ \boldsymbol{k}\in \{ \boldsymbol{k}|k<k_b \} } S_{\Omega , \boldsymbol{k}}$ for each $\Omega$ (see figure \ref{fig:pdw1}d). The numerically resolved $f_P(\Omega)$ is plotted in figure \ref{fig:f} at several different nonlinearity levels. We remark that these results are convergent in the sense that they are not sensitive to the further increase of the time window $T_w$, i.e., a longer time average. It is clear that under this level of average, $\int_\Omega f(\Omega)d\Omega\sim O(1)$ consistent with WTT (we do not expect $\int_\Omega f(\Omega)d\Omega$ to be exactly one unless we can numerically reach the kinetic limit). To quantify the profile of $f_P(\Omega)$, we use a least-square method to fit the data to a general functional form of $f_P(\Omega)=C/(\rho + \Omega ^{\beta})$, where $\rho$ is needed as a desingularisation factor for $f_P(0)$. These fittings, as shown in figure \ref{fig:f}, agree with all data points remarkably well, with $C=[1.06,0.982,0.875]$, $\rho=[6.176,3.44,2.26]$, and $\beta=[1.35,1.47,1.52]$ from high to low nonlinearity level. Instead of a $sinc$ function (which involves negative values that are not observed), the functional form of $f_P(\Omega)$ is somewhat closer to the WTT form of $\epsilon/(\Omega^2+\epsilon^2)$, but with different exponents $\beta$. This is probably why \citet{pan_understanding_2017} find that the kinetic equation employing this WTT form of generalized delta function produces physically reasonable results in terms of the spectral slope and energy flux for capillary waves. 

\section{Conclusion and Discussions}
In this paper, we numerically study the properties of inter-scale energy flux $P$ for wave turbulence in the context of the 2D MMT equation. Unlike previous evaluations of $P$ based on energy input or dissipation rate, our formulation of $P$ computes the exact instantaneous energy flux across arbitrary scale $k_b$ directly from the nonlinear terms in the MMT equation, and allows a quartet-level decomposition of $P$ into $P_\Omega$ according to the frequency mismatch $\Omega$ of the quartets. Our results show that the energy flux $P$ across any scale in the inertial range closely follows a Gaussian distribution, with their mean value $\overline{P}$ almost a constant for any $k_b$, and standard deviation $\sigma(P)$ increasing with $k_b$. In addition, values of $\sigma(P)$ are generally several orders of magnitude larger than $\overline{P}$, mainly due to the contributions to $\sigma(P)$ from quasi-resonances, i.e., $P_{\Omega>0}$. The decomposition of $P$ into $P_\Omega$ also allows an alternative but more direct measure of nonlinear broadening by quantitatively considering the contribution of quasi-resonances to the mean energy flux. We further study the scaling of spectral level $N$ with the energy flux and find that $N\sim \overline{P}^{1/3}$ (consistent with the kinetic scaling) at high nonlinearity and $N\sim \overline{P}^{1/2}$ (dynamic scaling) at low nonlinearity. The former and latter are due to the dominance of the quasi-resonant and exact-resonant contributions to $\overline{P}$, respectively. Finally, our numerical study on the wave-turbulence closure model shows that the fourth-order correlator is in disagreement with the description by the theoretical closure on a single quartet level. When considering the average over all quartets contributing to the inter-scale energy flux (over $O(10^9)$ quartets for each $\Omega$), more consistent behavior to the theoretical closure is observed, but with the broadening function exhibiting $1/\Omega^\beta$ (with $\beta$ between $1.3$ and $1.6$) different from the forms derived in WTT.

While this work sheds new light on the physics of wave turbulence, more unanswered questions about wave turbulence, especially regarding the closure model, are raised. The closure model is a subject that has received insufficient attention from a numerical perspective, especially in terms of analysis utilizing data generated directly from the primitive dynamic equation. The few exceptions to this \cite[e.g.][]{majda_one-dimensional_1997,annenkov_role_2006,sheffield_ensemble_2017} have not considered the detailed functional form and magnitude of $f(\Omega)$, as studied in this work. Therefore, the authors consider the primary importance of this work as to provide a methodology such that many open questions in wave turbulence can be directly studied using vast numerical data. Within the presented results, for example, the function $f_Q(\Omega)$ in figure \ref{fig:q1} is still sensitive to the time window $T_w$ used for averaging quantities in (\ref{eqn:closure}) for an individual quartet. It is not clear whether a convergent behavior (close to the quartet-averaged result $f_P(\Omega)$) can be found if a time average over an extremely long time window is performed. In addition, will the smooth behavior associated with function $Q_\Omega$ in figure \ref{fig:pdw1}e and \ref{fig:pdw1}f be preserved in different domain geometries (e.g., an irrational torus as in \citet{hrabski_energy_2021}) where the normalization factor $\mathcal{N}_\Omega$ substantially varies? Last but not least, are we able to explore the kinetic limit in numerical simulations by varying both domain size and nonlinearity level? If this is possible, will we find more consistent behavior for $f_F(\Omega)$ for families centered on different quartets, and will all these functions eventually converge to a delta function? 

These questions can be explored in greater detail with increasing computational resources, and it is certainly not unreasonable to think about studying wave turbulence with ``exascale computing'', an area under development for hydrodynamic turbulence \cite[e.g.][]{yeung_advancing_2020}. For wave turbulence, these resources may be better utilized in conjunction with an understanding of WTT, rather than simply boosting the resolution of simulations.  It might also be beneficial to revisit the one-dimensional MMT equation, where some of these questions can be investigated with a reduced computational cost but with the new techniques developed here. In doing this we may also hope to settle the dispute between the WTT and MMT closure models that have been haunting the field for more than 20 years.

 \section{Acknowledgements}
 This material is based upon work supported by the National Science Foundation Graduate Research Fellowship under Grant No. DGE 1841052. Any opinions, findings, and conclusions or recommendations expressed in this material are those of the authors and do not necessarily reflect the views of the National Science Foundation. This work used the Extreme Science and Engineering Discovery Environment (XSEDE), which is supported by National Science Foundation grant number ACI-1548562. Computation was performed on XSEDE Bridges-2 at the Pittsburgh Supercomputing Center through allocation PHY200041. This research was supported in part through computational resources and services provided by Advanced Research Computing (ARC), a division of Information and Technology Services (ITS) at the University of Michigan, Ann Arbor.

\bibliographystyle{jfm}
\bibliography{jfm_EP_STUDY}

\end{document}